\documentclass[conference]{IEEEtran}
\ifCLASSINFOpdf
\else
\fi
\usepackage{algorithm}
\usepackage[noend]{algpseudocode}
\usepackage{graphicx}
\usepackage[table,xcdraw]{xcolor}
\usepackage{multirow}
\usepackage[export]{adjustbox}

\begin{document}
%
\title{Evaluation of spatial trees \\ for simulation of biological tissue}

\author{\IEEEauthorblockN{Ilya Dmitrenok\IEEEauthorrefmark{1},
Viktor Drobnyy\IEEEauthorrefmark{1},
Leonard Johard and
Manuel Mazzara}
\IEEEauthorblockA{
Innopolis University\\
Innopolis, Russia}
\IEEEauthorblockA{\IEEEauthorrefmark{1} These authors contributed equally to this work.
}
}


%


\maketitle

\begin{abstract}
Spatial organization is a core challenge for all large agent-based models with local interactions. In biological tissue models, spatial search and reinsertion are frequently reported as the most expensive steps of the simulation. One of the main methods utilized in order to maintain both favourable algorithmic complexity and accuracy is spatial hierarchies. In this paper, we seek to clarify to which extent the choice of spatial tree affects performance, and also to identify which spatial tree families are optimal for such scenarios. We make use of a prototype of the new BioDynaMo tissue simulator for evaluating performances as well as for the implementation of the characteristics of several different trees. 

\end{abstract}


%
\IEEEpeerreviewmaketitle

\section{Introduction}
The high pace of neuroscientific research has led to a difficult problem in synthesizing the experimental results into effective new hypotheses. As an effort to understand how the the emergent properties of individual experiments neuroscientists has started to turn to simulators in increasing numbers. These simulators has proven to be one of the few effective tools that allows us study how individual known physical, biological or chemical principles give rise to complex emergent structures and behaviour in multicellular organism.

One of the most daunting challenges is to be able to study the interaction between principles that act across different time scales, such the relation between neuroplasticity and the dynamics of individual spike trains. The simulations need enough accuracy for short-term dynamics and be able to simulate periods of time sufficient for the long-scale dynamics to be evident. Such simulations will need advances in both computational efficiency and effective parallelization in order to be be practical as well as economically feasible.

\subsection{Spatial organization in simulators}
Cellular simulations have adapted a number of approaches with their respective advantages: lattice-based models and center-based models.  
A detailed overview is given in \cite{VanLiedekerke2015}, here we will birefly recall major aspects.

Lattice-based models put cells on a predefined lattice grid. This means they are limited in their ability to simulate dynamic processes, which is a significant drawback in the study of emergent behaviour. They also seem ill-suited to handle the various geometries involved in neurite growth.

Center-based models have similar run-times as lattice-based models, but are less restricted in its spatial structure. The main issue is that detailed cell anatomy requires significant computation time. The amount of cells practically simulated on a single computer generally stretches between a few thousands to a million, depending on detail level, hardware and simulated time range.

In center-based models, movement and neighborhood detection remains one of the main performance bottlenecks \cite{van2015simulating}. Performances in these bottlenecks are deeply intertwined with the spatial structure chosen for the simulation. 

\subsection{BioDynaMo}
The BioDynaMo project \cite{BreitwieserBMJK16} is developing a new general platform for computer simulations of biological tissue dynamics, with a brain development as a primary target. The platform should be executable on hybrid cloud computing systems, allowing for the efficient use of state-of-the-art computing technology.

The simulation will cover a range of cellular behaviours, including cell division, cell growth, DNA expression, electrophysiology, neurite extension, chemical gradients and mechanical forces. The main principle that effectively allows these simulations in a parallelized cloud structure is locality of interaction. Locality allows us to split the workload into spatial segments that need just limited communication between each other regarding the activity in border volumes. 

BioDynaMo uses a center-based model with explicit simulation of neurites, which puts among cented-based models with more detailed anatomy. Our simulation scale is similar to Cx3D \cite{zubler2009framework}, although we are adding cellular deformation and additional types of interaction between cells. This puts on a similar level of simulation detail as certain other simulators, perhaps most notably CellSys \cite{hoehme2010cell}. 

The aim of the BioDynaMo project is to push the limits of this simulation type with both highly efficient code and extensive parallelization on relatively cheap cloud-based hardware.

A central aspect in making such simulation possible is effective data structures. Since our simulation type is heavily exploiting the locality of the problem our most important data structure will be the spatial structures. These structures are important on two levels: First, we need a structure of the data for division among nodes and, secondly, we need a structure that is efficient for calculations in a single node.

\subsubsection{Distributed memory parallelization}
A full simulation of an adult human brain will require an estimated scale of 100.000 nodes. This requires effective spatial segmentation within the nodes, but also across nodes. Most importantly, the full model will be much to large in memory to be stored in each node and transferred across nodes in each time step of the simulation.

Any spatial partition will solve the memory limitation, but the limitations on bandwidth imposes limitation on geometry of our partitions, as the amount of border volume data in need of transmission will be approximately proportional to the surface area.

In combining bandwidth limitations, computational efficiency, ease of implementation and cost of load redistribution we an octree structure to be an appropriate method.

\subsubsection{Shared memory spatial organization}
Identifying an effective shared memory spatial organization is the main objective of this study. Since all threads have full access to the memory and the task itself is embarrassingly parallel, our main remaining concern is the algorithmic complexity. Although an octree is indeed a possiblity, like in the case of distributed memory parallelization, we can widen our search of candidates due to the lack of bandwidth limitations in shared memory conditions.

The runtime of tree data structures generally scales as $n log n$, unless we can exploit additional assumption on the data set. The main approaches in cented-based models are either Voronoi tesselation, e.g. Chaste \cite{mirams2013chaste}, or spatial trees, e.g. Timothy \cite{cytowski2014large}. 


Grids represent a simple collision detection method if we have an evenly distributed load. These are in many ways similar to the lattice-based models and share the same weakness at simultaneously simulating phenomena at different scales.

Recently, there has been an increasing interest in hashing functions. These allow an implicit representation of the spatial structure chosen and might speed up computations significantly \cite{choi2009linkless, lewiner2010fast}. The hash functions are an implicit representation of the spatial structure and mirror the grids or spatial trees.


\subsection{Problem formulation}
The algorithmic complexity of collision detection is data dependent and falls somewhere between $n$ and $n^2$n. This study seeks to identify suitable spatial hierarchies by performing empirical evaluations in a center-based cell simulation environment.

\section{Spatial trees}

There are several groups of spatial trees, each of which is related but not identical to some respective 1-dimensional equivalent. They can be divided into bounding volume hierarchies (threes of geometrical objects, OBB and AABB trees, K-dop, SSV, R trees) and spatial decompositions (BSP trees, k-d trees, MSP trees octrees, grids).

Whereas spatial decomposition is arranged around non-overlapping spatial regions, the bounding volume hierarchies consist of overlapping containers. 

\subsection{Performance considerations}
We are especially interested in two operations: 

\begin{itemize}
\item Neighborhood detection
\item Insertion
\end{itemize}
Neighborhood detection is to identify all potential neighbors within a radius r of the object. This is a preselection for the application of our local interaction rules. In collision detection terminology this is known as the broad phase detection and can be applied to improve the performance of any interaction with a limited range. For detection of actual collision we can then apply a stricter collision test, a so-called narrow phase, to each remaining preselected object.

Insertion is the update of the spatial tree as our objects change shape and move, or when new objects are added. Some tree structures can insert objects fast, but this fast insertion operation unbalances the tree and regular, and usually expensive, rebalancing operations have to be applied in order to optimie the tree.

Generally the theoretical bounds are the same and real performance depends heavily on the specifics of object movements and shapes. If we have $n$objects, most operations require between $n^2$ steps, representing pairwise comparisons between volumes, and n steps, representing an ideal constant time detection per object. Although many algorithm reach $n$ complexity under stricter assumptions, in the general case they reach an average complexity of $n log n$.

Certain geometries, such as convex hulls, allow a tighter fit to the objects and more selective neighborhood detection by a constant factor. This is usually counterbalanced by increased computation necessary in regular rebalancing of the trees.

Leonard, Ilya, Victor

\subsection{Octrees}
Octrees are one of the most intuitive tree structures and is considered easy to implement. It subdivides each node into eight octants, which keeps the cubical shape intact. This is useful when the aspect ratio of the node is important, as for minimizing contact surface in the cloud. 

The tree type is rather popular in the simulator community due to its ease of implementation and performance under frequent reinsertions. Some agent-based cell simulators using this approach is Timothy \cite{cytowski2015large}, Biocellion \cite{kang2014biocellion} and ...

\subsection{Kd tree}
K-d trees starts with a rectangle and iteratively cuts the rectangle with a splitting hyperplane orthogonal to one of its axes. Each subdivision becomes one of its children. 

In order to create a balanced three the easiest solution is simply to cycle between its axes between each cut, but there is uniquely defined way to construct a k-d tree.

The SEM++ simulator, utilizing the LAMMPS library, uses variants of this algorithm. In their variant each cut is done along its longest dimension. \cite{milde2014sem++, plimpton2007lammps}

\subsection{R-tree}
The R-tree is a spatial equivalent of the B-tree data structure. Being a bounded volume hierarchy and not a spatial tree, is substantially different from the other trees in this experimental comparison. 

An R-tree starts with a rectangular minimum bounding rectangle covering all children. Its children are in turn possible overlapping rectangle each covering all their children.

The insertion and deletion methods are weakly defined from such a structure and leaves many option open and is the main division between the various member algorithms of the R tree family. 

\subsubsection{Search}
Search in a regular R-tree consists of an iterative comparison of the search area with the bounding box. In BioDynaMo we use a bounding box covering our neighborhood and compare it with each children of the current node. Each child node that overlaps with our neighborhood box will be searched iteratively.

\subsubsection{Insertion}
Insertion consists of creating a bounding box of the object and identifying the appropriate subtree. In case there are multiple option, we insert into the tree that needs the least enlargement.

\section{Experiments}
The experimental part consists of comparing performance of the trees mentioned above. Each of them is a space-partitioning data structure for organizing points in the space. The similar structure of the  trees provided us with the possibility to use the same algorithm to perform a search query. Each of them has either an array of pointers to its children if it is an internal node or a container of objects if it is a leaf node.  All tree nodes have a bounding box that defines the positions of all objects of a corresponding subtree. The consistency of data structure and a unified searching algorithm allows us to measure performance of each tree within the equal conditions, concentrating on their space partitioning potential only.

Idea behind this algorithm is that we are making search between two trees and produce all pairs of close points where the first object is from tree $A$ and the second object is from tree $B$. Checking a leaf node will cause comparing all possible pairs. In the other cases, we are going deeper if and only if the distance between nodes stays smaller than required. To perform the search on one tree we have to call this function where $A$ = $B$ = tree. This code should be updated in order to manage cases where A and B is the same node. Algorithm is provided below.
\begin{algorithm}
\caption{Insertion Algorithm in octree and k-d}\label{alg:search}

\begin{algorithmic}[1]
\Procedure{insert}{point $p$, Type $object$}
\If{is\_leaf\_node}
\If{objects.size $>$ node\_capacity OR\\ \hfill max\_depth == 0}
\State objects.add($p$, $object$)
\Else
\State split() \Comment {method to split node space}
\State put($p$, $object$)
\EndIf
\Else
\State idx = get\_child\_id($p$)
\State children[idx].put($p$, $obj$)
\EndIf
\EndProcedure
\end{algorithmic}
\end{algorithm}
\begin{algorithm}
\caption{Search Algorithm}\label{alg:search}

\begin{algorithmic}[1]
\Procedure{search}{Tree $A$, Tree $B$, distance, result}
\If{$A$ is leaf and $B$ is leaf}
\For{each object $a$ in $A$}
        \For{each object $b$ in $B$}
        	\If{distance($a$,$b$) $\le$ distance}
            \State result.append(($a$,$b$))
        	\EndIf
      \EndFor
      \EndFor
\Else
	\If{$A$ is not leaf}
    \State aChildren = $A$.children
    \Else
    \State aChildren = [$A$]
    \EndIf
    \If{$B$ is not leaf}
    \State bChildren = $B$.children
    \Else
    \State bChildren = [$B$]
    \EndIf
    \For{each aChildren $a$}
    \For{each aChildren $b$}
    \If{minDistance($a$,$b$) $\le$ distance}
    \State search($a$, $b$, distance, result)
    \EndIf
    \EndFor
    \EndFor
\EndIf
\EndProcedure
\end{algorithmic}
\end{algorithm}

\subsection{Experimental simulator}

In order to compare trees, we have performed a series of tests on a random uniformly distributed data. Separate runs were performed with different parameters, like the maximum depth of the tree and maximum amount of objects in each node in case of an $octree$ and a $k-d$ tree or a degree in the case of an $r-tree$. Also, different ways of space partitioning were tested for a $k-d$ tree, including partition using a $median$, $center$ or $SAH$(surface area heuristics). The average time was calculated over 5 simulation runs. We separately kept truck of the time required for the two most important operations: insertion time and search time. 

Algorithms were implemented in C++ and all the experiments were run on Intel\textsuperscript{\textregistered} Core\texttrademark i5-4200U CPU @ 1.60GHz $\times$ 4. 


\section{Results}

Results of our experiment are presented in this section trough tables and charts: tables are providing the whole information on insertion time and searching time within two different distances. The purpose of the charts is to compare performance and to identify interesting trends. In order to improve readability and reduce the size of tables, the following abbreviations have been introduced:

\begin{itemize}
\item \textit{Tree(depth, capacity)} - shows type of the tree, its maximum depth and maximum capacity of the node, after which it will divide, in other words capacity is a division criteria. After reaching maximum depth tree node stops its dividing.
\item $MMAS$ - median multiple axis split. Space partitioning in such tree is based on median point of an axis. Axis is taken alternately, in xyz order. Sometimes denoted as simply a k-d tree.
\item $MSAS$ - median single axis split. Space partitioning in such tree is based on median point of an x axis only. 
\item $CS$ - center split. Space partitioning in such tree is based on center point of alternating xyz axis.
\item $SAHS$ - Surface Area Heuristics split. Surface Area Heuristics used as a splitting point.

\end{itemize}
\subsection{Impact of depth on insertion and search times wihin different distances}
\begin{figure}
\includegraphics[width=\linewidth, left]{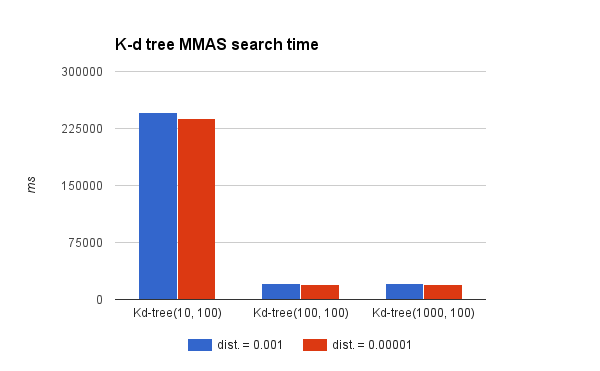}
\caption{Search time for k-d tree with MMAS space partitioning within distances 0.001 and 0.00001}
\label{fig:MMAS_st}
\end{figure}
\begin{figure}
\includegraphics[width=\linewidth, left]{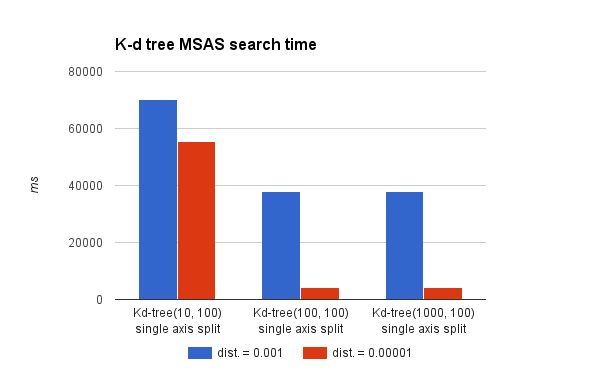}
\caption{Search time for k-d tree with MSAS space partitioning within distances 0.001 and 0.00001}
\label{fig:MSAS_st}
\end{figure}

As we can see from charts \ref{fig:MMAS_st} and \ref{fig:MSAS_st}, for the k-d tree, both with MMAS and MSAS, the trend is that the smaller the depth, the higher the searching time. It is much faster to search and prune nodes than objects within the node, so with the depth being too small, leaf nodes are overflowing, causing time increase. This illustrated the need to keep the maximum depth large enough to prevent leaf node overflow.

Also it is worth mentioning, that k-d tree with MMAS is much more insensitive to search distance than other tree types.
\subsection{Relation between insertion, search time and depth}
\begin{figure}
\includegraphics[width=\linewidth, left]{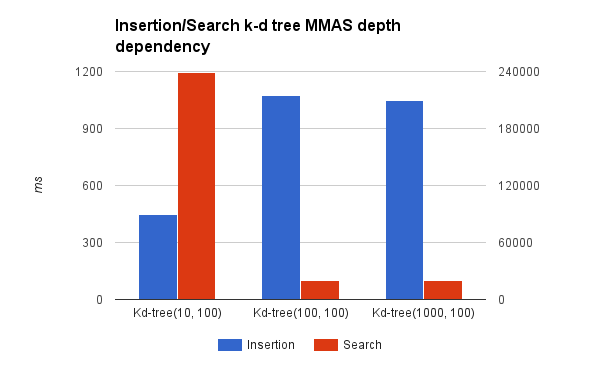}
\caption{Dependency of depth on search and insertion time for a k-d tree with MMAS space partitioning }
\label{fig:MMAS_depth}
\end{figure}
\begin{figure}
\includegraphics[width=\linewidth, left]{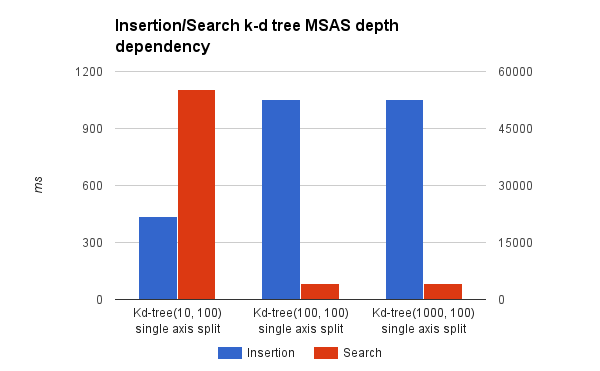}
\caption{Dependency of depth on search and insertion time for a k-d tree with MSAS space partitioning }
\label{fig:MSAS_depth}
\end{figure}

For the same reasons that were stated in the section $A$, the depth influences search time as follows: the smaller the depth, the higher the searching time, as can be observed from charts \ref{fig:MMAS_depth} and \ref{fig:MSAS_depth}. On the other hand, the smaller the depth, the faster the insertion time: as the insertion process progresses the depth of the tree grows. Consequently, it takes more and more time to reach the leaf node and split the node in case the tree has reached maximum capacity. With smaller depths this time will stop increasing sooner, although causing leaf node overflow begin. 

\subsection{Relation between insertion, search time and node capacity}
\begin{figure}
\includegraphics[width=\linewidth, left]{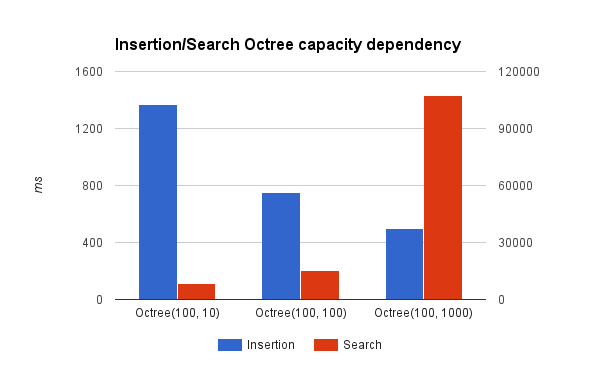}
\caption{Dependency of node capacity on search and insertion time for an octree}
\label{fig:octree_capacity}
\end{figure}

\begin{figure}
\includegraphics[width=\linewidth, left]{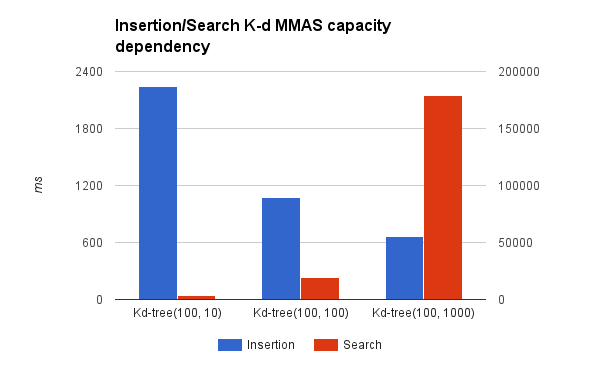}
\caption{Dependency of node capacity on search and insertion time for a k-d tree with MMAS space partitioning }
\label{fig:MMAS_capacity}
\end{figure}

\begin{figure}
\includegraphics[width=\linewidth, left]{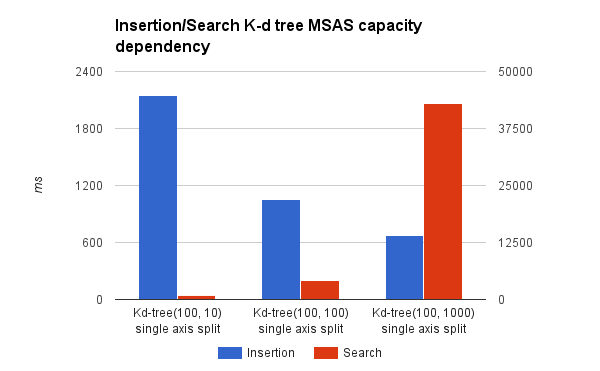}
\caption{Dependency of node capacity on search and insertion time for a k-d tree with MSAS space partitioning }
\label{fig:MSAS_capacity}
\end{figure}

This section is dedicated to comparison of how node capacity influence insertion time and search time.

As we can see from the charts \ref{fig:octree_capacity}, \ref{fig:MMAS_capacity}, and \ref{fig:MSAS_capacity}, higher the capacity, lower the search speed. With higher capacity the tree makes less splits, increasing insertion time. On the contrary, searching between objects of the leaf node costs much more than pruning and searching between the node.

\begin{table}[]
\centering
\caption{Insertion time, ms}
\label{insertion-time}
\begin{tabular}{|l|l|ll}
\hline
\rowcolor[HTML]{C0C0C0} 
\multicolumn{1}{|c|}{\cellcolor[HTML]{C0C0C0}} & \multicolumn{3}{c|}{\cellcolor[HTML]{C0C0C0}Number of cells} \\ \cline{2-4} 
\rowcolor[HTML]{EFEFEF} 
\multicolumn{1}{|c|}{\multirow{-2}{*}{\cellcolor[HTML]{C0C0C0}Tree(depth, capacity)}} & 10 000 & \multicolumn{1}{l|}{\cellcolor[HTML]{EFEFEF}100 000} & \multicolumn{1}{l|}{\cellcolor[HTML]{EFEFEF}1 000 000} \\ \hline
Octree(10, 10) & 15,38 & \multicolumn{1}{l|}{182,46} & \multicolumn{1}{l|}{1 503,01} \\ \hline
Octree(100, 10) & 14,96 & \multicolumn{1}{l|}{177,95} & \multicolumn{1}{l|}{1 371,88} \\ \hline
Octree(1000, 10) & 14,88 & \multicolumn{1}{l|}{170,80} & \multicolumn{1}{l|}{1 349,98} \\ \hline
Octree(10, 100) & 6,60 & \multicolumn{1}{l|}{62,85} & \multicolumn{1}{l|}{801,71} \\ \hline
Octree(100, 100) & 6,17 & \multicolumn{1}{l|}{61,56} & \multicolumn{1}{l|}{751,49} \\ \hline
Octree(1000, 100) & 5,69 & \multicolumn{1}{l|}{61,44} & \multicolumn{1}{l|}{741,70} \\ \hline
Octree(10, 1000) & 3,63 & \multicolumn{1}{l|}{50,35} & \multicolumn{1}{l|}{490,99} \\ \hline
Octree(100, 1000) & 3,66 & \multicolumn{1}{l|}{57,58} & \multicolumn{1}{l|}{497,98} \\ \hline
\textbf{Octree(1000, 1000)} & 
\textbf{3,65} & \multicolumn{1}{l|}{\textbf{44,99}} & \multicolumn{1}{l|}{\textbf{484,47}} \\ \hline
\textbf{k-d tree(10, 10) MMAS} & \textbf{14,08} & \multicolumn{1}{l|}{\textbf{77,32}} & \multicolumn{1}{l|}{\textbf{436,55}} \\ \hline
k-d tree(100, 10) MMAS & 20,92 & \multicolumn{1}{l|}{274,88} & \multicolumn{1}{l|}{2 240,25} \\ \hline
k-d tree(1000, 10) MMAS & 21,21 & \multicolumn{1}{l|}{268,25} & \multicolumn{1}{l|}{2 194,12} \\ \hline
k-d tree(10, 100) MMAS & 6,66 & \multicolumn{1}{l|}{81,99} & \multicolumn{1}{l|}{445,47} \\ \hline
k-d tree(100, 100) MMAS & 6,56 & \multicolumn{1}{l|}{88,24} & \multicolumn{1}{l|}{1 073,71} \\ \hline
k-d tree(1000, 100) MMAS & 6,55 & \multicolumn{1}{l|}{81,66} & \multicolumn{1}{l|}{1 048,03} \\ \hline
k-d tree(10, 1000) MMAS & 4,77 & \multicolumn{1}{l|}{53,00} & \multicolumn{1}{l|}{573,81} \\ \hline
k-d tree(100, 1000) MMAS & 4,77 & \multicolumn{1}{l|}{54,56} & \multicolumn{1}{l|}{664,63} \\ \hline
k-d tree(1000, 1000) MMAS & 4,74 & \multicolumn{1}{l|}{55,03} & \multicolumn{1}{l|}{665,25} \\ \hline
\textbf{k-d tree(10, 10) MSAS} & \textbf{14,15} & \multicolumn{1}{l|}{\textbf{76,63}} & \multicolumn{1}{l|}{\textbf{392,06}} \\ \hline
k-d tree(100, 10) MSAS & 22,14 & \multicolumn{1}{l|}{254,92} & \multicolumn{1}{l|}{2 147,12} \\ \hline
k-d tree(1000, 10) MSAS & 21,62 & \multicolumn{1}{l|}{245,70} & \multicolumn{1}{l|}{2 175,72} \\ \hline
k-d tree(10, 100) MSAS & 6,65 & \multicolumn{1}{l|}{71,92} & \multicolumn{1}{l|}{436,19} \\ \hline
k-d tree(100, 100) MSAS & 6,64 & \multicolumn{1}{l|}{83,03} & \multicolumn{1}{l|}{1 053,41} \\ \hline
k-d tree(1000, 100) MSAS & 6,69 & \multicolumn{1}{l|}{82,51} & \multicolumn{1}{l|}{1 052,75} \\ \hline
k-d tree(10, 1000) MSAS & 4,94 & \multicolumn{1}{l|}{54,16} & \multicolumn{1}{l|}{582,30} \\ \hline
k-d tree(100, 1000) MSAS & 4,93 & \multicolumn{1}{l|}{54,73} & \multicolumn{1}{l|}{669,05} \\ \hline
k-d tree(1000, 1000) MSAS & 4,91 & \multicolumn{1}{l|}{54,80} & \multicolumn{1}{l|}{672,99} \\ \hline
k-d tree(10, 100) CS & 3,09 &  &  \\ \cline{1-2}
k-d tree(100, 100) CS & 23,85 &  &  \\ \cline{1-2}
k-d tree(1000, 100) CS & 344,41 &  &  \\ \cline{1-2}
k-d tree(10, 1000) CS & 4,08 &  &  \\ \cline{1-2}
k-d tree(100, 1000) CS & 32,85 &  &  \\ \cline{1-2}
k-d tree(1000, 1000) CS & 381,81 &  &  \\ \cline{1-2}
k-d tree(10, 100) SAHS & 3,86 &  &  \\ \cline{1-2}
k-d tree(100, 100) SAHS & 27,38 &  &  \\ \cline{1-2}
k-d tree(1000, 100) SAHS & 364,26 &  &  \\ \cline{1-2}
k-d tree(10, 1000) SAHS & 4,43 &  &  \\ \cline{1-2}
k-d tree(100, 1000) SAHS & 35,25 &  &  \\ \cline{1-2}
k-d tree(1000, 1000) SAHS & 419,87 &  &  \\ \cline{1-2}
R-tree(5) & 47,39 &  &  \\ \cline{1-2}
R-tree(25) & 64,73 &  &  \\ \cline{1-2}
R-tree(125) & 65,80 &  &  \\ \cline{1-2}
\end{tabular}
\par
\bigskip
Table shows the total time of insertion operation for different amount of elements. Trees with different maximum depth and node capacity are considered. Best performance for each tree  is highlited in bold.
\end{table}

\begin{table}[]
\centering
\caption{Searching time in the distance 0,001, ms}
\label{searching-time-0.001}
\begin{tabular}{|l|l|ll}
\hline
\rowcolor[HTML]{C0C0C0} 
\multicolumn{1}{|c|}{\cellcolor[HTML]{C0C0C0}} & \multicolumn{3}{c|}{\cellcolor[HTML]{C0C0C0}Number of objects} \\ \cline{2-4} 
\rowcolor[HTML]{EFEFEF} 
\multicolumn{1}{|c|}{\multirow{-2}{*}{\cellcolor[HTML]{C0C0C0}Tree(depth, capacity)}} & 10000 & \multicolumn{1}{l|}{\cellcolor[HTML]{EFEFEF}100000} & \multicolumn{1}{l|}{\cellcolor[HTML]{EFEFEF}1000000} \\ \hline
Octree(10, 10) & 131,79 & \multicolumn{1}{l|}{1136,53} & \multicolumn{1}{l|}{8151,38} \\ \hline
Octree(100, 10) & 127,23 & \multicolumn{1}{l|}{1160,41} & \multicolumn{1}{l|}{8069,68} \\ \hline
\textbf{Octree(1000, 10)} & \textbf{115,46} & \multicolumn{1}{l|}{\textbf{963,16}} & \multicolumn{1}{l|}{\textbf{8069,13}} \\ \hline
Octree(10, 100) & 83,8 & \multicolumn{1}{l|}{1175} & \multicolumn{1}{l|}{15363,76} \\ \hline
Octree(100, 100) & 94,78 & \multicolumn{1}{l|}{1191,66} & \multicolumn{1}{l|}{15385,8} \\ \hline
Octree(1000, 100) & 84,86 & \multicolumn{1}{l|}{1174,96} & \multicolumn{1}{l|}{15370,73} \\ \hline
Octree(10, 1000) & 455,35 & \multicolumn{1}{l|}{7748,83} & \multicolumn{1}{l|}{107569,86} \\ \hline
Octree(100, 1000) & 444,02 & \multicolumn{1}{l|}{7466,19} & \multicolumn{1}{l|}{107504,09} \\ \hline
Octree(1000, 1000) & 448,29 & \multicolumn{1}{l|}{7451,38} & \multicolumn{1}{l|}{107217,69} \\ \hline
k-d tree(10, 10) MMAS & 71,86 & \multicolumn{1}{l|}{4343,45} & \multicolumn{1}{l|}{424274,44} \\ \hline
k-d tree(100, 10) MMAS & 38,87 & \multicolumn{1}{l|}{356,5} & \multicolumn{1}{l|}{3447,13} \\ \hline
\textbf{k-d tree(1000, 10) MMAS} & \textbf{37,03} & \multicolumn{1}{l|}{\textbf{347,72}} & \multicolumn{1}{l|}{\textbf{3411,07}} \\ \hline
k-d tree(10, 100) MMAS & 152 & \multicolumn{1}{l|}{2548,28} & \multicolumn{1}{l|}{245501,51} \\ \hline
k-d tree(100, 100) MMAS & 152,72 & \multicolumn{1}{l|}{1826,13} & \multicolumn{1}{l|}{20647,72} \\ \hline
k-d tree(1000, 100) MMAS & 151,08 & \multicolumn{1}{l|}{1832,5} & \multicolumn{1}{l|}{20635,61} \\ \hline
k-d tree(10, 1000) MMAS & 784,94 & \multicolumn{1}{l|}{15755,85} & \multicolumn{1}{l|}{249615,4} \\ \hline
k-d tree(100, 1000) MMAS & 792,57 & \multicolumn{1}{l|}{15699,33} & \multicolumn{1}{l|}{194723,31} \\ \hline
k-d tree(1000, 1000) MMAS & 782,70 & \multicolumn{1}{l|}{15652,81} & \multicolumn{1}{l|}{194537,06} \\ \hline
k-d tree(10, 10) MSAS & 21,54 & \multicolumn{1}{l|}{1471,95} & \multicolumn{1}{l|}{148399,95} \\ \hline
k-d tree(100, 10) MSAS & 12,63 & \multicolumn{1}{l|}{622,44} & \multicolumn{1}{l|}{46982,82} \\ \hline
k-d tree(1000, 10) MSAS & 12,78 & \multicolumn{1}{l|}{598,91} & \multicolumn{1}{l|}{46661,78} \\ \hline
k-d tree(10, 100) MSAS & 42,10 & \multicolumn{1}{l|}{738,19} & \multicolumn{1}{l|}{70331,8} \\ \hline
\textbf{k-d tree(100, 100) MSAS} & \textbf{42,06} & \multicolumn{1}{l|}{\textbf{655,64}} & \multicolumn{1}{l|}{\textbf{37905,61}} \\ \hline
k-d tree(1000, 100) MSAS & 43,04 & \multicolumn{1}{l|}{656,33} & \multicolumn{1}{l|}{37951,93} \\ \hline
k-d tree(10, 1000) MSAS & 331,85 & \multicolumn{1}{l|}{4434,3} & \multicolumn{1}{l|}{74047,06} \\ \hline
k-d tree(100, 1000) MSAS & 328,90 & \multicolumn{1}{l|}{4352,96} & \multicolumn{1}{l|}{68095,3} \\ \hline
k-d tree(1000, 1000) MSAS & 331,62 & \multicolumn{1}{l|}{4354,4} & \multicolumn{1}{l|}{68045,4} \\ \hline
k-d tree(10, 100) CS & 1844,87 &  &  \\ \cline{1-2}
k-d tree(100, 100) CS & 1974,1 &  &  \\ \cline{1-2}
k-d tree(1000, 100) CS & 2295,9 &  &  \\ \cline{1-2}
k-d tree(10, 1000) CS & 1874,93 &  &  \\ \cline{1-2}
k-d tree(100, 1000) CS & 1878,83 &  &  \\ \cline{1-2}
k-d tree(1000, 1000) CS & 2009,10 &  &  \\ \cline{1-2}
k-d tree(10, 100) SAHS & 1889 &  &  \\ \cline{1-2}
k-d tree(100, 100) SAHS & 1879,69 &  &  \\ \cline{1-2}
k-d tree(1000, 100) SAHS & 2206,6 &  &  \\ \cline{1-2}
k-d tree(10, 1000) SAHS & 1898,88 &  &  \\ \cline{1-2}
k-d tree(100, 1000) SAHS & 1954,66 &  &  \\ \cline{1-2}
k-d tree(1000, 1000) SAHS & 2058,93 &  &  \\ \cline{1-2}
R-tree(5) & 1493,98 &  &  \\ \cline{1-2}
R-tree(25) & 1918,96 &  &  \\ \cline{1-2}
R-tree(125) & 1863,96 &  &  \\ \cline{1-2}
\end{tabular}
\par
\bigskip
Table shows total time of search within distance 0.001 for different amount of elements has being inserted. Trees with different maximum depth and node capacity are considered. Best performance for each tree is highlited in bold.
\end{table}

\begin{table}[]
\centering
\caption{Searching time in the distance 0,00001, ms}
\label{searching-time-0.00001}
\begin{tabular}{|l|l|ll}
\hline
\rowcolor[HTML]{C0C0C0} 
\multicolumn{1}{|c|}{\cellcolor[HTML]{C0C0C0}} & \multicolumn{3}{c|}{\cellcolor[HTML]{C0C0C0}Number of objects} \\ \cline{2-4} 
\rowcolor[HTML]{EFEFEF} 
\multicolumn{1}{|c|}{\multirow{-2}{*}{\cellcolor[HTML]{C0C0C0}Tree(depth, capacity)}} & 10000 & \multicolumn{1}{l|}{\cellcolor[HTML]{EFEFEF}100000} & \multicolumn{1}{l|}{\cellcolor[HTML]{EFEFEF}1000000} \\ \hline
Octree(10, 10) & 127,47 & \multicolumn{1}{l|}{1 156,65} & \multicolumn{1}{l|}{8 105,95} \\ \hline
Octree(100, 10) & 117,35 & \multicolumn{1}{l|}{1 052,75} & \multicolumn{1}{l|}{8 180,85} \\ \hline
\textbf{Octree(1000, 10)} & \textbf{118,55} & \multicolumn{1}{l|}{\textbf{965,34}} & \multicolumn{1}{l|}{\textbf{8 075,78}} \\ \hline
Octree(10, 100) & 92,51 & \multicolumn{1}{l|}{1 184,20} & \multicolumn{1}{l|}{15 401,08} \\ \hline
Octree(100, 100) & 86,53 & \multicolumn{1}{l|}{1 175,68} & \multicolumn{1}{l|}{15 358,40} \\ \hline
Octree(1000, 100) & 91,34 & \multicolumn{1}{l|}{1 186,09} & \multicolumn{1}{l|}{15 381,49} \\ \hline
Octree(10, 1000) & 446,38 & \multicolumn{1}{l|}{7 459,64} & \multicolumn{1}{l|}{107 221,02} \\ \hline
Octree(100, 1000) & 444,94 & \multicolumn{1}{l|}{7 449,14} & \multicolumn{1}{l|}{107 243,58} \\ \hline
Octree(1000, 1000) & 455,29 & \multicolumn{1}{l|}{7 465,83} & \multicolumn{1}{l|}{107 363,20} \\ \hline
k-d tree(10, 10) MMAS & 71,19 & \multicolumn{1}{l|}{4 259,27} & \multicolumn{1}{l|}{413 708,11} \\ \hline
k-d tree(100, 10) MMAS & 37,34 & \multicolumn{1}{l|}{349,50} & \multicolumn{1}{l|}{3 226,78} \\ \hline
\textbf{k-d tree(1000, 10) MMAS} & \textbf{35,73} & \multicolumn{1}{l|}{\textbf{328,20}} & \multicolumn{1}{l|}{\textbf{3 207,74}} \\ \hline
k-d tree(10, 100) MMAS & 148,36 & \multicolumn{1}{l|}{2 473,96} & \multicolumn{1}{l|}{238 809,07} \\ \hline
k-d tree(100, 100) MMAS & 147,61 & \multicolumn{1}{l|}{1 783,90} & \multicolumn{1}{l|}{19 534,04} \\ \hline
k-d tree(1000, 100) MMAS & 148,40 & \multicolumn{1}{l|}{1 783,17} & \multicolumn{1}{l|}{19 476,90} \\ \hline
k-d tree(10, 1000) MMAS & 771,35 & \multicolumn{1}{l|}{15 073,20} & \multicolumn{1}{l|}{230 123,63} \\ \hline
k-d tree(100, 1000) MMAS & 775,92 & \multicolumn{1}{l|}{15 079,19} & \multicolumn{1}{l|}{179 362,17} \\ \hline
k-d tree(1000, 1000) MMAS & 770,34 & \multicolumn{1}{l|}{15 108,05} & \multicolumn{1}{l|}{179 354,74} \\ \hline
k-d tree(10, 10) MSAS & 19,47 & \multicolumn{1}{l|}{1 350,45} & \multicolumn{1}{l|}{129 661,86} \\ \hline
k-d tree(100, 10) MSAS & 8,39 & \multicolumn{1}{l|}{81,15} & \multicolumn{1}{l|}{974,00} \\ \hline
\textbf{k-d tree(1000, 10) MSAS} & \textbf{8,68} & \multicolumn{1}{l|}{\textbf{80,76}} & \multicolumn{1}{l|}{\textbf{973,72}} \\ \hline
k-d tree(10, 100) MSAS & 42,39 & \multicolumn{1}{l|}{590,59} & \multicolumn{1}{l|}{55 297,02} \\ \hline
k-d tree(100, 100) MSAS & 43,21 & \multicolumn{1}{l|}{414,49} & \multicolumn{1}{l|}{4 112,81} \\ \hline
k-d tree(1000, 100) MSAS & 42,81 & \multicolumn{1}{l|}{418,61} & \multicolumn{1}{l|}{4 128,74} \\ \hline
k-d tree(10, 1000) MSAS & 330,87 & \multicolumn{1}{l|}{4 359,53} & \multicolumn{1}{l|}{55 010,07} \\ \hline
k-d tree(100, 1000) MSAS & 329,58 & \multicolumn{1}{l|}{4 356,88} & \multicolumn{1}{l|}{42 990,69} \\ \hline
k-d tree(1000, 1000) MSAS & 331,17 & \multicolumn{1}{l|}{4 359,36} & \multicolumn{1}{l|}{42 957,82} \\ \hline

k-d tree(10, 100) CS & 1 844,87 &  &  \\ \cline{1-2}
k-d tree(100, 100) CS & 1 918,36 &  &  \\ \cline{1-2}
k-d tree(1000, 100) CS & 2 226,06 &  &  \\ \cline{1-2}
k-d tree(10, 1000) CS & 1 863,67 &  &  \\ \cline{1-2}
k-d tree(100, 1000) CS & 1 856,19 &  &  \\ \cline{1-2}
k-d tree(1000, 1000) CS & 1 971,19 &  &  \\ \cline{1-2}
k-d tree(10, 100) SAHS & 1 869,63 &  &  \\ \cline{1-2}
k-d tree(100, 100) SAHS & 1 899,31 &  &  \\ \cline{1-2}
k-d tree(1000, 100) SAHS & 2 217,93 &  &  \\ \cline{1-2}
k-d tree(10, 1000) SAHS & 1 929,73 &  &  \\ \cline{1-2}
k-d tree(100, 1000) SAHS & 1 900,82 &  &  \\ \cline{1-2}
k-d tree(1000, 1000) SAHS & 2 020,87 &  &  \\ \cline{1-2}
R-tree(5) & 1 490,22 &  &  \\ \cline{1-2}
R-tree(25) & 1 897,58 &  &  \\ \cline{1-2}
R-tree(125) & 1 868,17 &  &  \\ \cline{1-2}
\end{tabular}
\par
\bigskip
  Table shows total time of search within distance 0.00001 for different amount of elements has being inserted. Trees with different maximum depth and node capacity are considered. Best performance for each tree is highlited in bold. 
\end{table}

\begin{table}[]
\centering
\caption{Total insertion + searching time for 1M objects within different distances}
\label{insertion+search}
\begin{tabular}{|l|l|l|}
\hline
\rowcolor[HTML]{C0C0C0} 
\cellcolor[HTML]{C0C0C0} & \multicolumn{2}{c|}{\cellcolor[HTML]{C0C0C0}Time, ms} \\ \cline{2-3} 
\rowcolor[HTML]{EFEFEF} 
\cellcolor[HTML]{C0C0C0} & \multicolumn{2}{c|}{\cellcolor[HTML]{EFEFEF}Total insertion+search} \\ \cline{2-3} 
\rowcolor[HTML]{EFEFEF} 
\multirow{-3}{*}{\cellcolor[HTML]{C0C0C0}Tree(depth, capacity)} & Distance 0.001 & Distance 0.00001 \\ \hline
Octree(10, 10) & 9 654,39 & 9 608,96 \\ \hline
Octree(100, 10) & 9 441,57 & 9 552,73 \\ \hline
Octree(1000, 10) & 9 419,11 & 9 425,76 \\ \hline
Octree(10, 100) & 16 165,47 & 16 202,79 \\ \hline
Octree(100, 100) & 16 137,38 & 16 109,89 \\ \hline
Octree(1000, 100) & 16 112,44 & 16 123,19 \\ \hline
Octree(10, 1000) & 108 060,85 & 107 712,01 \\ \hline
Octree(100, 1000) & 108 002,08 & 107 741,56 \\ \hline
Octree(1000, 1000) & 107 702,16 & 107 847,66 \\ \hline
Kd-tree(10, 10) MMAS & 424 710,99 & 414 144,66 \\ \hline
Kd-tree(100, 10) MMAS & 5 687,38 & 5 467,02 \\ \hline
\textbf{Kd-tree(1000, 10) MMAS} & \textbf{5 605,20} & 5 401,86 \\ \hline
Kd-tree(10, 100) MMAS & 245 946,99 & 239 254,54 \\ \hline
Kd-tree(100, 100) MMAS & 21 721,44 & 20 607,75 \\ \hline
Kd-tree(1000, 100) MMAS & 21 683,64 & 20 524,93 \\ \hline
Kd-tree(10, 1000) MMAS & 250 189,29 & 230 697,44 \\ \hline
Kd-tree(100, 1000) MMAS & 195 387,95 & 180 026,80 \\ \hline
Kd-tree(1000, 1000) MMAS & 195 202,31 & 180 019,99 \\ \hline
Kd-tree(10, 10) MSAS & 148 792,02 & 130 053,92 \\ \hline
\textbf{Kd-tree(100, 10) MSAS} & 49 129,94 & \textbf{3 119,11} \\ \hline
Kd-tree(1000, 10) MSAS & 48 837,50 & 3 149,44 \\ \hline
Kd-tree(10, 100) MSAS & 70 767,99 & 55 733,21 \\ \hline
Kd-tree(100, 100) MSAS & 38 959,02 & 5 166,22 \\ \hline
Kd-tree(1000, 100) MSAS & 39 004,68 & 5 181,49 \\ \hline
Kd-tree(10, 1000) MSAS & 74 629,36 & 55 592,37 \\ \hline
Kd-tree(100, 1000) MSAS & 68 764,35 & 43 659,74 \\ \hline
Kd-tree(1000, 1000) MSAS & 68 718,45 & 43 630,81 \\ \hline
\end{tabular}
\par
\bigskip
  Table shows total time of insertion 1.000.000 objects in the tree and then performing search operation within different distances. Trees with different maximum depth and node capacity are considered. Best performance for each distance is highlited in bold. 
\end{table}

\section{Discussion}
Even though the insertion time is a relevant parameter, it was found to be small in comparison with the search time on average. From our point of view it means it is much more expedient to decrease searching time, even if the cost is rising insertion times.

Although we expected insertion times to be a bottleneck for R-trees, they surprisingly performed substantially worse also on search. The heuristic balancing proved ineffective on our data set.

Overall, our preliminary results supports the use of spatial hierarchies over bounding volume hierarchies and motivates the current use of octrees in center-based cellular simulations. Since the cuts of spatial division are able to separate data in distributed memory setups, such data division seems preferable at all levels of cloud-based neighborhood detection.

In future work we will proceed with the study of several more advanced derivative tree structures and more diverse experiments, like bulk load insertion and tree performance on different object space distributions. We will also be able benchmark our trees in a new and more feature complete prototype of the BioDynaMo simulator, allowing is to measure performance of more complex simulation runs ported from Cx3D \cite{zubler2013simulating}.

Further, we are exploring development of new and more dedicated spatial structures that can further exploit the locality and limited positioning differences between updates. 

Consequent developments in broad phase neighborhood detection has potential impact not only on BioDynaMo center-based cellular models, but also in related areas of collision detection and computer graphics.

\section{Conclusion}
Neighborhood detection is widely reported to be the most expensive algorithm in center-based cellular simulation. We have performed simulations on random data and found that in this case the bottleneck is search time. The higher complexity and increased insertion times of R trees did not results in better balanced time, but rather a significantly lower performance in this setting.

Overall our results support the use of k-d trees for simulation, while octrees have intermediate performance.

The experimental results also demonstrate the balancing act implied in choosing suitable parameter values as well as the magnitude of their performance impact. The tables approximately identify the critical points in such performance optimization.

When the experimental results of the BioDynaMo prototype are published, the results published here will function as reference values.

\section*{Acknowledgements}
We would like to thank Innopolis University for logistic and financial support. Our gratitude goes to colleagues at IU who participated in discussions and seminars and actively supported this research.






%


\bibliographystyle{IEEEtran}
\bibliography{IEEEabrv,mainbib}

\begin{thebibliography}{10}
\providecommand{\url}[1]{#1}
\csname url@samestyle\endcsname
\providecommand{\newblock}{\relax}
\providecommand{\bibinfo}[2]{#2}
\providecommand{\BIBentrySTDinterwordspacing}{\spaceskip=0pt\relax}
\providecommand{\BIBentryALTinterwordstretchfactor}{4}
\providecommand{\BIBentryALTinterwordspacing}{\spaceskip=\fontdimen2\font plus
\BIBentryALTinterwordstretchfactor\fontdimen3\font minus
  \fontdimen4\font\relax}
\providecommand{\BIBforeignlanguage}[2]{{%
\expandafter\ifx\csname l@#1\endcsname\relax
\typeout{** WARNING: IEEEtran.bst: No hyphenation pattern has been}%
\typeout{** loaded for the language `#1'. Using the pattern for}%
\typeout{** the default language instead.}%
\else
\language=\csname l@#1\endcsname
\fi
#2}}
\providecommand{\BIBdecl}{\relax}
\BIBdecl

\bibitem{VanLiedekerke2015}
\BIBentryALTinterwordspacing
P.~Van~Liedekerke, M.~M. Palm, N.~Jagiella, and D.~Drasdo, ``Simulating tissue
  mechanics with agent-based models: concepts, perspectives and some novel
  results,'' \emph{Computational Particle Mechanics}, vol.~2, no.~4, pp.
  401--444, 2015. [Online]. Available:
  \url{http://dx.doi.org/10.1007/s40571-015-0082-3}
\BIBentrySTDinterwordspacing

\bibitem{van2015simulating}
P.~Van~Liedekerke, M.~Palm, N.~Jagiella, and D.~Drasdo, ``Simulating tissue
  mechanics with agent-based models: concepts, perspectives and some novel
  results,'' \emph{Computational Particle Mechanics}, vol.~2, no.~4, pp.
  401--444, 2015.

\bibitem{BreitwieserBMJK16}
L.~Breitwieser, R.~Bauer, A.~D. Meglio, L.~Johard, M.~Kaiser, M.~Manca,
  M.~Mazzara, F.~Rademakers, and M.~Talanov, ``The biodynamo project: Creating
  a platform for large-scale reproducible biological simulations,'' \emph{4th
  Workshop on Sustainable Software for Science: Practice and Experiences
  (WSSSPE4)}, 2016.

\bibitem{zubler2009framework}
F.~Zubler and R.~Douglas, ``A framework for modeling the growth and development
  of neurons and networks,'' \emph{Frontiers in computational neuroscience},
  vol.~3, p.~25, 2009.

\bibitem{hoehme2010cell}
S.~Hoehme and D.~Drasdo, ``A cell-based simulation software for multi-cellular
  systems,'' \emph{Bioinformatics}, vol.~26, no.~20, pp. 2641--2642, 2010.

\bibitem{mirams2013chaste}
G.~R. Mirams, C.~J. Arthurs, M.~O. Bernabeu, R.~Bordas, J.~Cooper, A.~Corrias,
  Y.~Davit, S.-J. Dunn, A.~G. Fletcher, D.~G. Harvey \emph{et~al.}, ``Chaste:
  an open source c++ library for computational physiology and biology,''
  \emph{PLoS Comput Biol}, vol.~9, no.~3, p. e1002970, 2013.

\bibitem{cytowski2014large}
M.~Cytowski and Z.~Szymanska, ``Large-scale parallel simulations of 3d cell
  colony dynamics,'' \emph{Computing in Science \& Engineering}, vol.~16,
  no.~5, pp. 86--95, 2014.

\bibitem{choi2009linkless}
M.~G. Choi, E.~Ju, J.-W. Chang, J.~Lee, and Y.~J. Kim, ``Linkless octree using
  multi-level perfect hashing,'' in \emph{Computer Graphics Forum}, vol.~28,
  no.~7.\hskip 1em plus 0.5em minus 0.4em\relax Wiley Online Library, 2009, pp.
  1773--1780.

\bibitem{lewiner2010fast}
T.~Lewiner, V.~Mello, A.~Peixoto, S.~Pesco, and H.~Lopes, ``Fast generation of
  pointerless octree duals,'' in \emph{Computer Graphics Forum}, vol.~29,
  no.~5.\hskip 1em plus 0.5em minus 0.4em\relax Wiley Online Library, 2010, pp.
  1661--1669.

\bibitem{cytowski2015large}
M.~Cytowski and Z.~Szymanska, ``Large-scale parallel simulations of 3d cell
  colony dynamics: The cellular environment,'' \emph{Computing in Science \&
  Engineering}, vol.~17, no.~5, pp. 44--48, 2015.

\bibitem{kang2014biocellion}
S.~Kang, S.~Kahan, J.~McDermott, N.~Flann, and I.~Shmulevich, ``Biocellion:
  accelerating computer simulation of multicellular biological system models,''
  \emph{Bioinformatics}, vol.~30, no.~21, pp. 3101--3108, 2014.

\bibitem{milde2014sem++}
F.~Milde, G.~Tauriello, H.~Haberkern, and P.~Koumoutsakos, ``Sem++: a particle
  model of cellular growth, signaling and migration,'' \emph{Computational
  Particle Mechanics}, vol.~1, no.~2, pp. 211--227, 2014.

\bibitem{plimpton2007lammps}
S.~Plimpton, P.~Crozier, and A.~Thompson, ``Lammps-large-scale atomic/molecular
  massively parallel simulator,'' \emph{Sandia National Laboratories}, vol.~18,
  2007.

\bibitem{zubler2013simulating}
F.~Zubler, A.~Hauri, S.~Pfister, R.~Bauer, J.~C. Anderson, A.~M. Whatley, and
  R.~J. Douglas, ``Simulating cortical development as a self constructing
  process: a novel multi-scale approach combining molecular and physical
  aspects,'' \emph{PLoS Comput Biol}, vol.~9, no.~8, p. e1003173, 2013.

\end{thebibliography}

\end{document}